\documentclass{article}
\usepackage[T1]{fontenc} 
\usepackage[utf8]{inputenc} 
\usepackage{ismir,amsmath,cite,url}
\usepackage{graphicx}
\usepackage{color}
\usepackage{csquotes}
\usepackage{amssymb}
\usepackage{pifont}
\usepackage[compact]{titlesec}
\titlespacing{\section}{1.5pt}{1.5ex}{0ex}
\titlespacing{\subsection}{0pt}{0.5ex}{0ex}
\titlespacing{\subsubsection}{0pt}{0.5ex}{0ex}
\UseRawInputEncoding

\usepackage{lineno}

\title{\enquote*{1e0a}: A COMPUTATIONAL APPROACH TO RHYTHM TRAINING}

\oneauthor
{Names should be omitted for double-blind reviewing}
{Affiliations should be omitted for double-blind reviewing}

\twoauthors
  {Noel Alben} {Electronics and Communication, \\ PESIT - BSC, Bangalore}
 {Ranjani H.G.} {GAIA, \\ Ericsson Research, Bangalore}

\begin{document}

	\maketitle
	\begin{abstract}
		
		We present a computational assessment system that promotes the learning of basic rhythmic patterns. The system is capable of generating multiple rhythmic patterns with increasing complexity within various cycle lengths. For a generated rhythm pattern the performance assessment of the learner is carried out through the statistical deviations calculated from the onset detection and temporal assessment of a learner's performance. This is compared with the generated pattern, and their performance accuracy forms the feedback to the learner. The system proceeds to generate a new pattern of increased complexity when performance assessment results are within certain error bounds. The system thus mimics a learner-teacher relationship as the learner progresses in their feedback-based learning. The choice of progression within a cycle for each pattern is determined by a predefined complexity metric. This metric is based on a coded element model for the perceptual processing of sequential stimuli. The model earlier proposed for a sequence of tones and non-tones, is now used for onsets and silences. This system is developed into a web-based application and provides accessibility for learning purposes. Analysis of the performance assessments shows that the complexity metric is indicative of the perceptual processing of rhythm patterns and can be used for rhythm learning. 
		
	\end{abstract}

	\section{Introduction}\label{sec:introduction}
	Learning to play a musical instrument is a complex process
	that requires the acquisition of many interlinked skills such as tone, intonation, note accuracy, timing, rhythm precision, clear articulation, and dynamic variation [1]. An acknowledged important skill is the understanding of rhythm, timing, and rhythm precision. 
	Rhythm is defined to be the description and understanding of
	the duration and durational patterns of musical notes [2]. The
	word rhythm is used to refer to all of the temporal aspects of
	a musical work [3].\\
	Two main contexts within which formal musical learning takes place are the
	classroom, through playing and interacting with a teacher,
	and at home, practicing alone [1]. From the perspective of
	a drum kit or percussive instrument, a typical classroom
	setting involves the teacher and a learner. The teacher decides what the learner must practice and the teacher's feedback helps them achieve accuracy in timing and quality in performance. The learner's practicing habits at home aid in reaching performance goals. A software
	application cannot substitute a music teacher but it can
	be a compliment in the learning process, especially in the
	performance assessment of daily practice at home [1].
	
	\subsection{Related Literature}
	Most assessment methods are either the assessment of learning, used to measure competence, or the assessment for learning used to enhance
	learning [4-6]. The rapid development of music technology over the last few decades has dramatically changed the way that people interact with music today. In [7], the authors explore the use of Music Information Retrieval (MIR) techniques in music education and their integration in learning software. They outline Song2See as a representative example of a music learning system developed within the MIR community. Songs2See is a music game developed based on pitch detection, sound separation, music transcription, interface development, and audio analysis technologies [8].
	Despite advancements in these sound analysis technologies, Eremenko V, Morsi A, et al. in [1], revealed that the performance assessment tools present in these applications are still not reliable and effective enough to support the strict requirements of a professional music education context. In [1], they review some of the work done in the practice of music assessment and present a proof of concept for a complete assessment system for supporting guitar exercises. Furthermore, they identify the challenges that should be addressed to further advance these assessment technologies and their useful integration into professional learning contexts. Particularly those, that will help learners to plan, monitor, and evaluate their learning progression. They also highlight the importance and requirement of presenting the assessment results to a learner in a way that promotes learning [1]. Similarly, Lerch, A, et al. in [9] surveyed the field of Music Performance Analysis (MPA) from the perspective of performance assessment and its ability to provide insights and interactive feedback by analyzing and assessing the audio of practice sessions. Furthermore, they highlight the necessity for individual assessment and curriculum in the context of a music education tool. The work of Chih-Wei Wu and Alexander Lerch in [10], presents an unsupervised feature learning approach to derive features for assessing percussive instrument performances amongst a large data of audition performances, mimicking a judge in this scenario. Specifically, they proposed using a histogram-based input representation to sparse coding to allow the sparse coding to take advantage of temporal rhythmic information. Various (commercially available) solutions exist specifically for rhythm education and performance assessment. Examples for rhythm tutoring applications are Perfect Ear [11], Rhythm Trainer [12], and Complete Rhythm Trainer [13]. These form a class of mobile applications that help learners practice and assess their rhythm skills by following a set curriculum presented in each of these applications and touchscreen-based feedback for performance assessment. Other apps such as SmartMusic [14], Yousician [15], Music Prodigy [16], and SingStar [17] are available for music education purposes however, they do not have a dedicated rhythm training section. Furthermore, none of the previous works address musical rhythm complexity measures as a feature to aid in a learner's progression. Evaluating various rhythm complexity measures based on how accurately they reflect human-based measures were studied by Eric Thul in his thesis [3]. Although his results suggest that none of the measures accurately reflect the difficulty humans have performing or listening to rhythm, the measures do accurately reflect how humans recognize a rhythm's structure.

	\subsection{Our Contribution}
	In this work, we consider the individual assessment of learning a particular rhythmic pattern at a fixed tempo. We present the proof of concept for a system that will assess a learner's audio recording and provide feedback to the learner on their performance. We will only focus on assessing the learner's temporal proficiency, keeping the performance parameter categories of tempo, dynamics, pitch, and timbre fixed. Furthermore, we focus on the progression of a learner within a certain cycle length. We do so with the introduction of new patterns of increased complexity based on a pre-defined complexity metric that is similar to a teacher who decides what the learner must practice. 
	\\
	This manuscript is organized as follows, we first define
	the terms, notations, and conventions to be held throughout
	that will help assess the performance of an individual. That
	is followed by the proposed system for the assessment
	where we go over the subsequent blocks that make up
	the assessment system and understand the coded element model used to define the complexity metric. Finally, we discuss the learner results and accuracy of this system through our web application platform, and conclude with future enhancements and improvements to the system.
	\section{Background}
	\subsection{Definitions}{
		\begin{figure}[h!]
			\centerline{\framebox{
					\includegraphics[width=7.5cm]{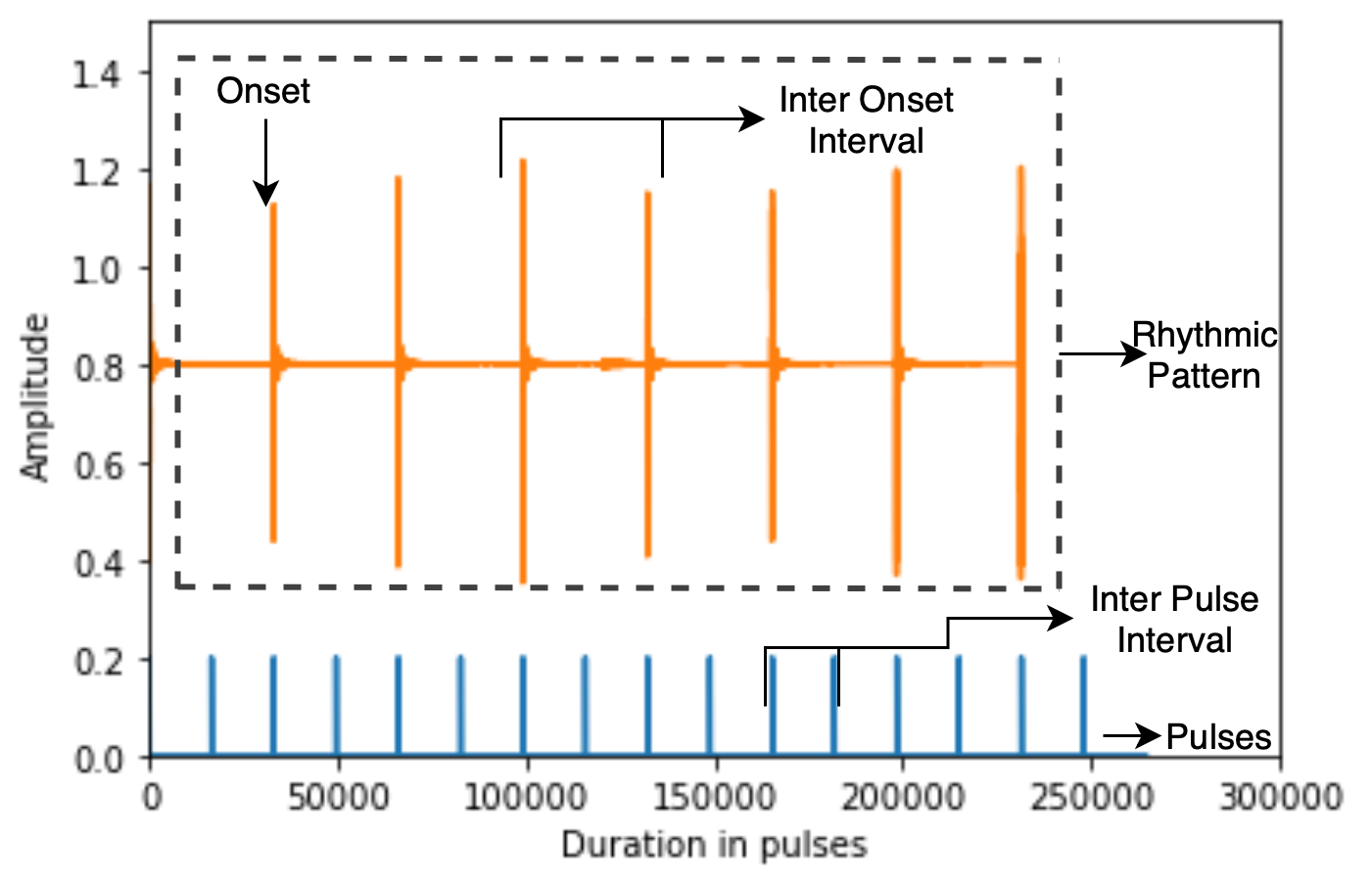}}}
			\caption{Diagrammatic illustration of the terms used in this
				article}
			\label{fig:Def1}
		\end{figure}
		We first provide a brief background of some of the terms used to define a rhythmic pattern.\\
		In this work we consider the fundamental and individual time unit for a rhythmic pattern to be a \textit{pulse}.
		Pulse is defined to be  \enquote{one of a series of regularly recurring,
			precisely equivalent stimuli} [18]. The duration of a rhythmic pattern is composed of a number of pulses. Tempo refers to the rate at which the musical pulses are generated and it also controls the number of pulses for a given duration. For our proposed system the total duration of a rhythmic pattern is determined by the total number of samples present in the audio signal. The uniform number of samples or duration, between two subsequent pulses, is defined to be the inter-pulse interval (IPI). The IPI depends on the sampling frequency and the tempo used for the rhythmic pattern. Figure \ref{fig:Def1} shows a duration of 300000 samples with pulses generated at a fixed Inter Pulse Interval (IPI), determined by a tempo of 160 Pulses Per Minute (PPM). \\
		To generate a rhythmic pattern, we consider instantaneous attacks on the marked time units (pulses) i.e., striking a drum, clapping, or tapping. The instantaneous attacks are defined as onsets. The duration that is marked between two onsets is defined as
		the inter-onset interval (IOI). The IPI, onsets, and the IOI together create a rhythmic pattern as shown in figure \ref{fig:Def1}.
		\\
		A cycle determines the number of pulses beyond which the rhythmic pattern begins to repeat itself. The cycle represents a structure for the pulses, and this structure
		is realised by the rhythmic pattern. For example, we use 7 to indicate a repeating rhythmic pattern of 7 pulses. Rhythm patterns can be composed in any predetermined cycle number, for this article we will only be following cycles of 4,3,5,7, and 16.

	}
	\subsection{Representation}
\begin{figure}[h!]

		\begin{minipage}[b]{0.48\linewidth}
			\centering
			\centerline{\includegraphics[width=4.8cm]{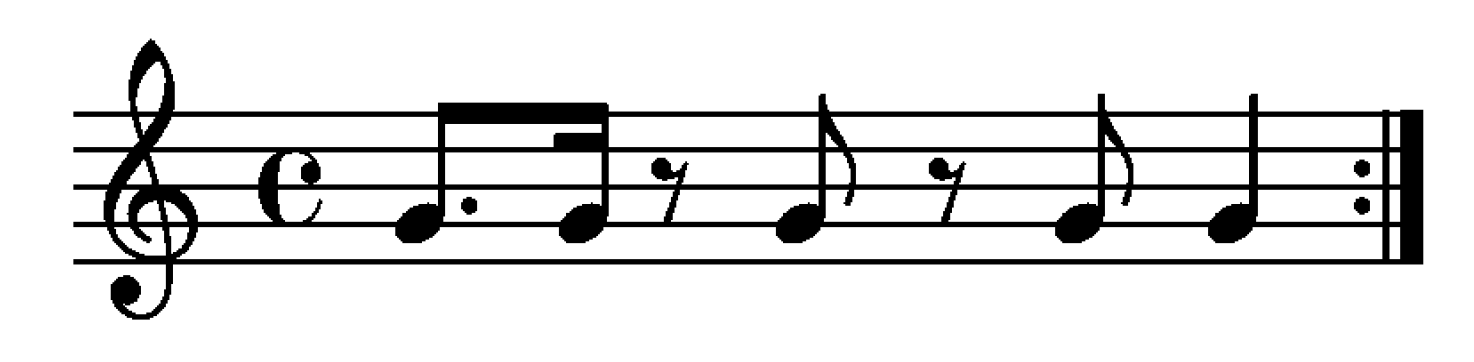}}
			\centerline{(a)}\medskip
		\end{minipage}
		\hspace{0.15cm}
		\begin{minipage}[b]{0.48\linewidth}
			\centering
			\centerline{\includegraphics[width=3.5cm, height = 0.75cm]{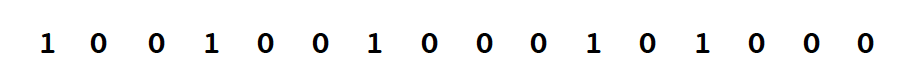}}
			\centerline{(b)}\medskip
		\end{minipage}
		\caption{ (a) The clave son rhythm in music (stave) notation. (b) The clave son rhythm in binary notation.}
		\label{fig:bin}
	\end{figure}

	We will now introduce the rhythm pattern representation used in the proposed system with the help of a famous Afro-Cuban rhythm, the clave son, shown in figure \ref{fig:bin} (a).
	
	Figure. \ref{fig:bin} (b) shows the clave son rhythm as a binary string, a notation that describes the position of the onsets and silences. This notation is used in the domain of computer science [19]. The onsets are marked by a \enquote{1} and the silent pulses are marked by a \enquote{0}. This representation enables faster computation and integration of rhythm patterns into our system. This representation of a rhythm pattern henceforth will be referred to as the binary representation. 
	\subsection{ Naming 1e0a}
	The name \enquote*{1e0a} for the proposed system is derived from a mixture of the binary representation with its 1's and 0's and a western rhythm counting technique used by a large majority of music educators, called the Harr system. The system that Harr used assigned a set of numbers and syllables. For a cycle of length 4, numbers are assigned to onsets that occur at the beginning of the rhythmic pattern's duration and its repeats. Syllables are assigned to those onsets that fall everywhere else. The syllables remain
	consistent for all the pulses with only the number for each repeat changing. For example, the rhythm $[1 1 1 1 1 1 1 1]$ would be read, \enquote{1 e \& a 2 e \& a} [20]. Thereby, giving us the name \textbf{\enquote*{1e0a}}.
	
	\section{Proposed System}
	\begin{figure}[htb]
		\centering
		\centerline{\includegraphics[width=7.2cm, height = 5cm]{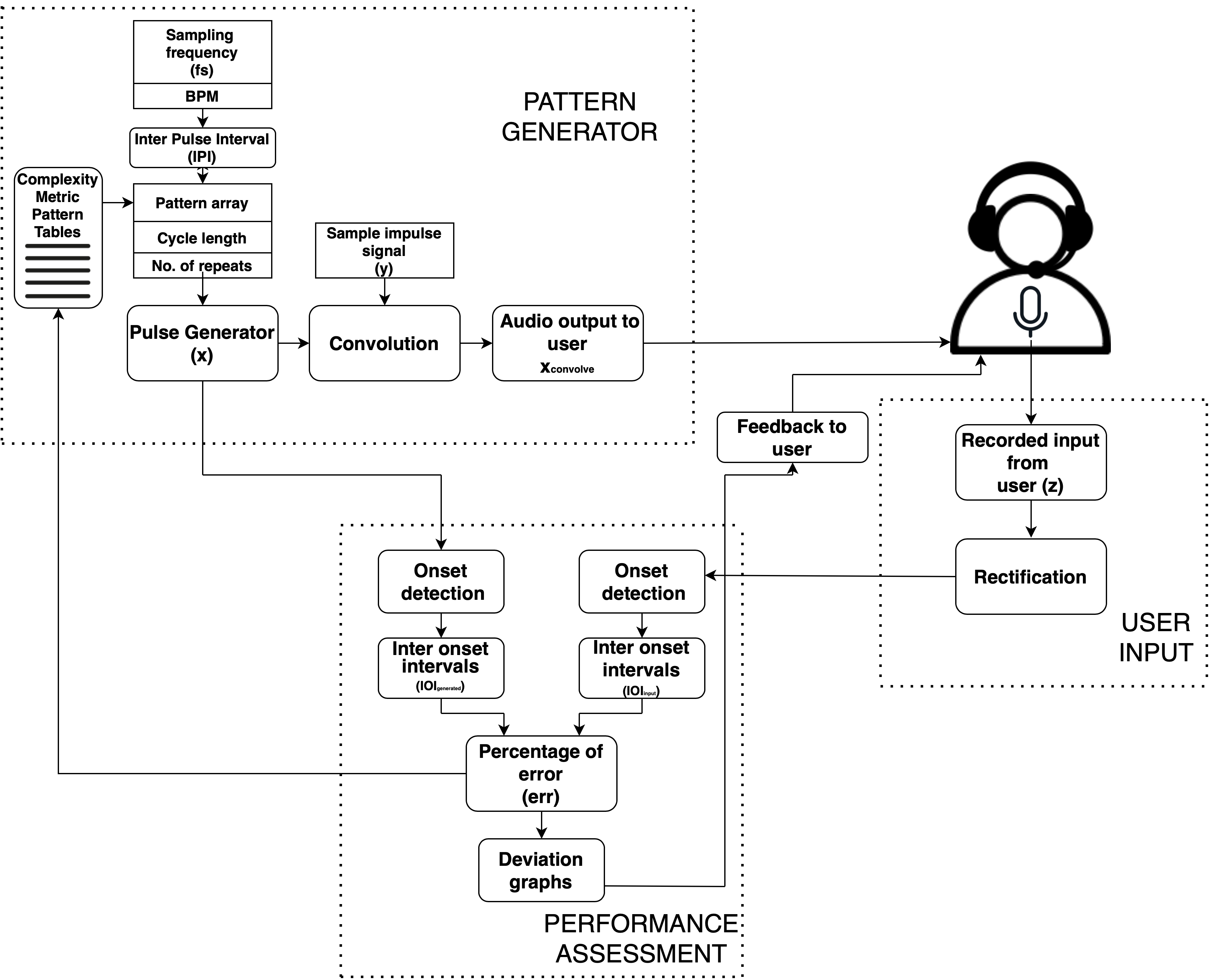}}
		\caption{Block diagram of the proposed system for \enquote*{1e0a.}}
		\label{fig:flow}
	\end{figure}
	Figure \ref{fig:flow} shows the block diagram of our proposed
	system for rhythm training and progression in learning. This system consists of two distinct
	blocks: (1) The rhythm pattern generator with the complexity metric, and, (2) the performance assessment of the learner input with feedback.
	
	\subsection{Pattern Generator}
	\subsubsection{Pattern Generation}\label{gen}
	The pattern generator as shown in figure 5 begins with a pulse generator function. This function takes input parameters that determine the cycle length, pattern array,
	tempo (PPM), and the total number of repeats,
	required to generate a collection of pulses whose inter-pulse
	interval (IPI) is determined by the equation \ref{sfeq}. \\

	\begin{equation}
		IPI = (60/PPM)/(1/fs) 
		\label{sfeq}
	\end{equation}
	
\begin{figure}[h!]

		\begin{minipage}[b]{0.48\linewidth}
			\centering
			\centerline{\includegraphics[width=3.8cm]{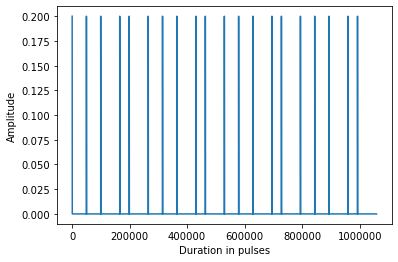}}
			\centerline{(a)}\medskip
		\end{minipage}
		\hspace{0.15cm}
		\begin{minipage}[b]{0.48\linewidth}
			\centering
			\centerline{\includegraphics[width=3.8cm]{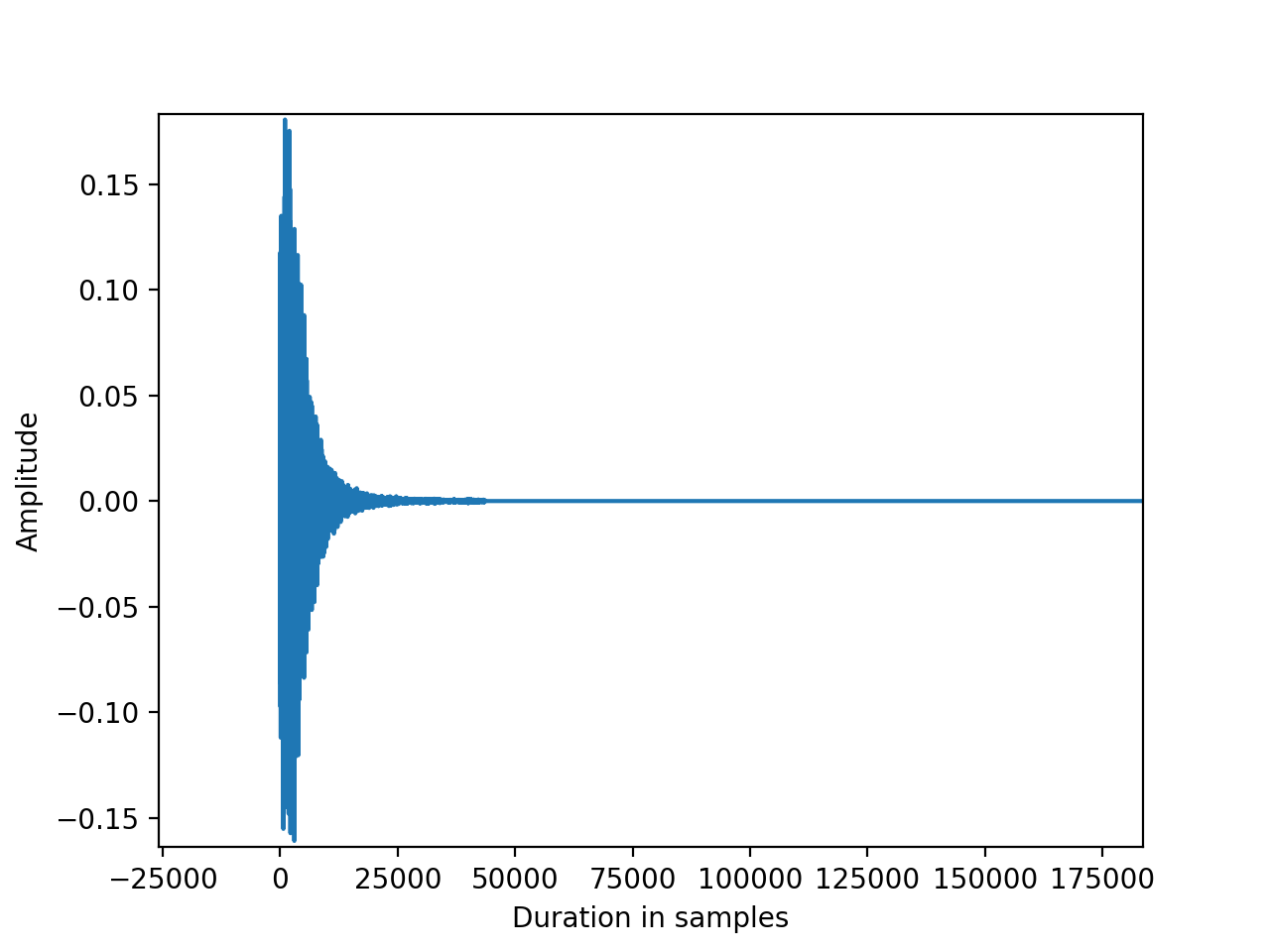}}
			\centerline{(b)}\medskip
		\end{minipage}
		\begin{minipage}[b]{\linewidth}		\centerline{\includegraphics[width=3.8cm]{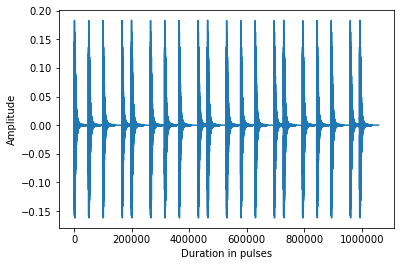}}
			\centerline{(c)}
		\end{minipage}
		\caption{Pattern Generator plots. (a) Pulse Generator Output; (b) Hi-Hat Drum sample impulse signal; (c) Convolution result of (a) and (b).}
		\label{fig:ex1}
		
	\end{figure}

$PPM$ represents the pulses per minute of the generated collection of pulses, $fs$ represents the sampling frequency and $IPI$ represents the fixed inter-pulse interval.
	\\
	The pattern array in binary representation determines the location of onsets and the number of silences to be incorporated into the duration of pulses. Furthermore, we propose the selection of pattern arrays for generating subsequent sequences using complexity metrics detailed in section \ref{complexity}.
	\\
	For the clave son: [1 0 0 1 0 0 1 0 0 0 1 0 1 0 0 0] , at 160 $ PPM $, with 4 repeats, at an $fs$ of 44100. The function generates an array of pulses as shown in figure \ref{fig:ex1}(a). The fixed inter-pulse interval for 160 $PPM$ as calculated by equation \ref{sfeq} is 16539 samples.

	The next stage of the pattern generator block adds tonal
	quality to the output of the pulse generator. We achieve this
	by convolving the pulse generator output with an appropriate
	drum sample impulse signal. The drum sample impulse signal is a short recording of an onset played using a particular drum sound. The drum sound selected must have an instantaneous attack and a short decay
	as shown in figure \ref{fig:ex1}(b).\\
	The convolution process requires the loaded sample impulse
	to have the same $fs$ as the pulse generator output. For
	a pulse generator output and drum impulse, figure \ref{fig:ex1}(c)
	shows the output of the convolution function. We make
	use of the Discrete Fourier Transform convolution property [21]. This allows our result to have the same number of samples
	as the pulse generator output. We use the Fast Fourier
	Transform (FFT) [21] algorithm to achieve this result.\\
	The resultant clave son rhythm from the pattern generator block, at 160 $ PPM $ and 4 cycles, is an audio signal whose plot is shown in figure \ref{fig:ex1}(c).

	\subsubsection{Complexity Metric}\label{complexity}
	
	Various measures of musical rhythm complexity have been studied in [3]. Rhythm complexity measures were divided into:
	\\ \textbf{Rhythm syncopation} measures that use metrical weights determined from the metrical hierarchy to find syncopated onsets in a rhythm [19,22]. 
	\\ \textbf{Pattern matching} measures which chop up a rhythm as determined by the levels of the metrical hierarchy and search each sub-rhythm for patterns that indicate the complexity [23,24].
	\\ \textbf{Distance measures}, which measure how far away a rhythm is from a more simple rhythm composed of beats [19,25,26].
	\\ Finally, mathematical measures which are based upon \textbf{Shannon's information entropy} [27,28]. \textbf{Shannon's information entropy} uncertainty equations are given in equations \ref{eqn1},\ref{eqn2},\ref{eqn3}.
	\\ We propose to use the rhythmic complexity measure that calculates information entropy in the sense of \textbf{Shannon's entropy} for a particular rhythm pattern. Vitz and Todd in [28], present a method that can be used to calculate the complexity of cyclic binary patterns. Their work defined that the complexity measure of a certain periodic or repeating pattern is the sum of the total amount of uncertainty evaluated in processing the stimulus sequence. In [28], they achieved this by proposing a set of rules (axioms) describing a perceptual coding process. As the method implies an emphasis on coded elements, it is referred to as a coded element processing system (CEPS) [3]. 
	\\We consider this model for our application due to its algorithmic approach to measure rhythm complexity, thus making it viable for coding, as against rhythm syncopation models in [19,21]. CEPS also has an improved ability to distinguish rhythm patterns within a cycle based on complexity as compared to pattern matching [23,24] and distance measures [19,25,26]. 
	\\
	The axioms in [28], proposed originally for tones and non-tones, break down sequences into code levels. In our work, we extend it to rhythmic patterns, composed of onsets and silences. Each code level successfully codes the pattern from single elements to larger elements which continues until the entire pattern is constructed.
	To calculate the complexity of the rhythmic pattern, we measure two uncertainty values at each Code Level in the algorithm as proposed in [28]. Uncertainty in this case means information entropy. The uncertainty values are the maximum uncertainty, $H_{max}$, and the joint uncertainty, $H_{joint}$ represented in equations \ref{hmax} and \ref{Hjoint} respectively.

	\begin{equation}
		H_{max}(X,Y,Z,W) = H(X) +H(Y) +H(Z) + H(W)
		\label{hmax}
	\end{equation}
	\begin{align}
		H_{joint}(X,Y,Z,W) = &H(X)+H(Y|X)+H(Z|X,Y) \notag\\                         &+H(W| W,Y,Z)
		\label{Hjoint}
	\end{align}
	where,
	\begin{equation}
		H(X) = - \sum_{x\in{X}}^{} p(x)log_2p(x)
		\label{eqn1}
	\end{equation}
	\begin{equation}
		H(X,Y) = - \sum_{x\in{X}}^{}\sum_{y\in{Y}}^{}p(x)p(x,y) log_2p(x,y)
		\label{eqn2}
	\end{equation}
	\begin{equation}
		H(X|Y) = - \sum_{x\in{X}}^{}\sum_{y\in{Y}}^{}p(y,x) log_2p(y/x)
		\label{eqn3}
	\end{equation}
	
	Let the value $H^k$ be the sum of $H^k_{max}$ and $H^k_{joint}$ at Code Level k. The total sum of H at each Code Level represents the complexity of the pattern. This value is termed $H_{code}$.
	\begin{equation}
		H_{code} = \sum_{k=1}^{n}H^k = \sum_{k=1}^{n}H_{max}^k+H_{joint}^k
	\end{equation}
	The complexity measure for $[1 0 0 1 0 0 1 0 0 0 1 0 1 0 0 0 ]$ is calculated to be \textbf{20.5538}. Similarly, we were able to calculate complexity measures for all the combinations of patterns within the cycle lengths of 4,3,5, and 7. Upon calculating the complexity measures, we arrange patterns according to their increasing order of complexity as shown in $sheets$\footnote{\url{{https://docs.google.com/spreadsheets/d/e/2PACX-1vTSDSqil264H6-M2kLdQUuD6VYnRqYPQePhKl8C_STD1j-54rFdhbLag-ep_15ggiHpo8-DOpJWyQ-O/pubhtml}}}. We observed that for patterns that are a shifted version of another, where 1's (onsets) replace the 0's (silences), and vice versa, the complexity measures are identical. For example, $[1 1 0 0]$ and $[0 0 1 1]$ have a complexity measure of 5.0. To tackle this, we considered those patterns that begin with a 0 as off-beat patterns and assumed that they present a higher challenge to learners. Hence, we incremented a constant value to the complexity measures for all those patterns that begin with a 0 (silence). The constant value within a cycle was determined by the highest complexity measure for the patterns that begin with 1 (onsets), for that cycle length. This can be observed in $sheets^{1}$.\\
	Furthermore, this complexity metric does not account for the complexity that arises with changes in tempo. Therefore in this work, we only assess the performances at a constant $PPM$.
	
	\subsection{Performance Assessment}\label{assess}
	Once the learner listens to the generated pattern their performance is recorded and loaded onto the system. We rectify the recorded audio signal $(z)$ with a threshold, $z_{threshold} = z_{max}/4$, to improve the results of the onset detection function and it also accounts for low magnitude noise in the live recording environment. 
	
	\begin{figure}[h!]

		\begin{minipage}[b]{0.48\linewidth}
			\centering
			\centerline{\includegraphics[width=4.0cm]{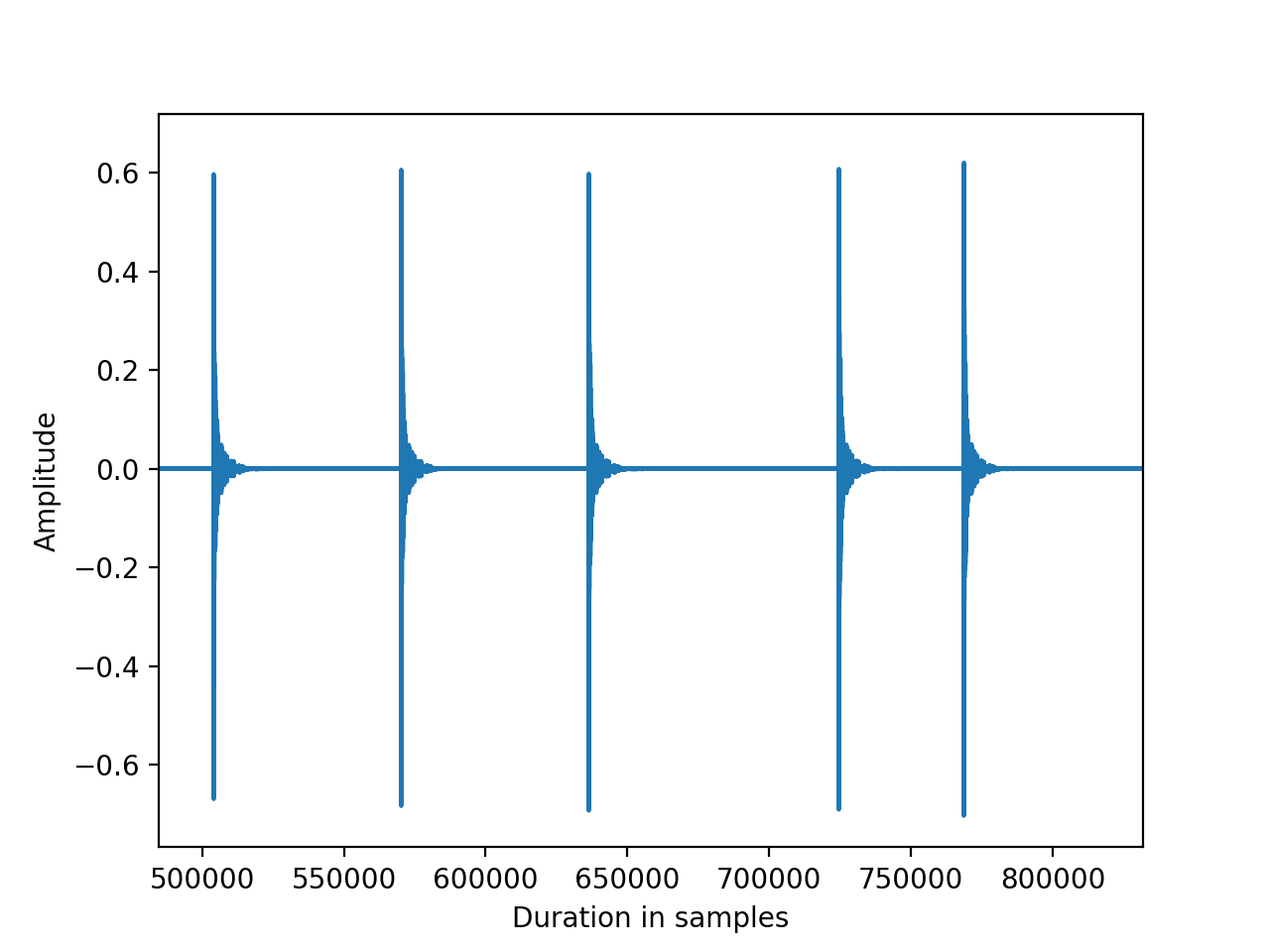}}
			\centerline{(a) Recorded input.}\medskip
		\end{minipage}
		\hfill
		\begin{minipage}[b]{0.48\linewidth}
			\centering
			\centerline{\includegraphics[width=4.0cm]{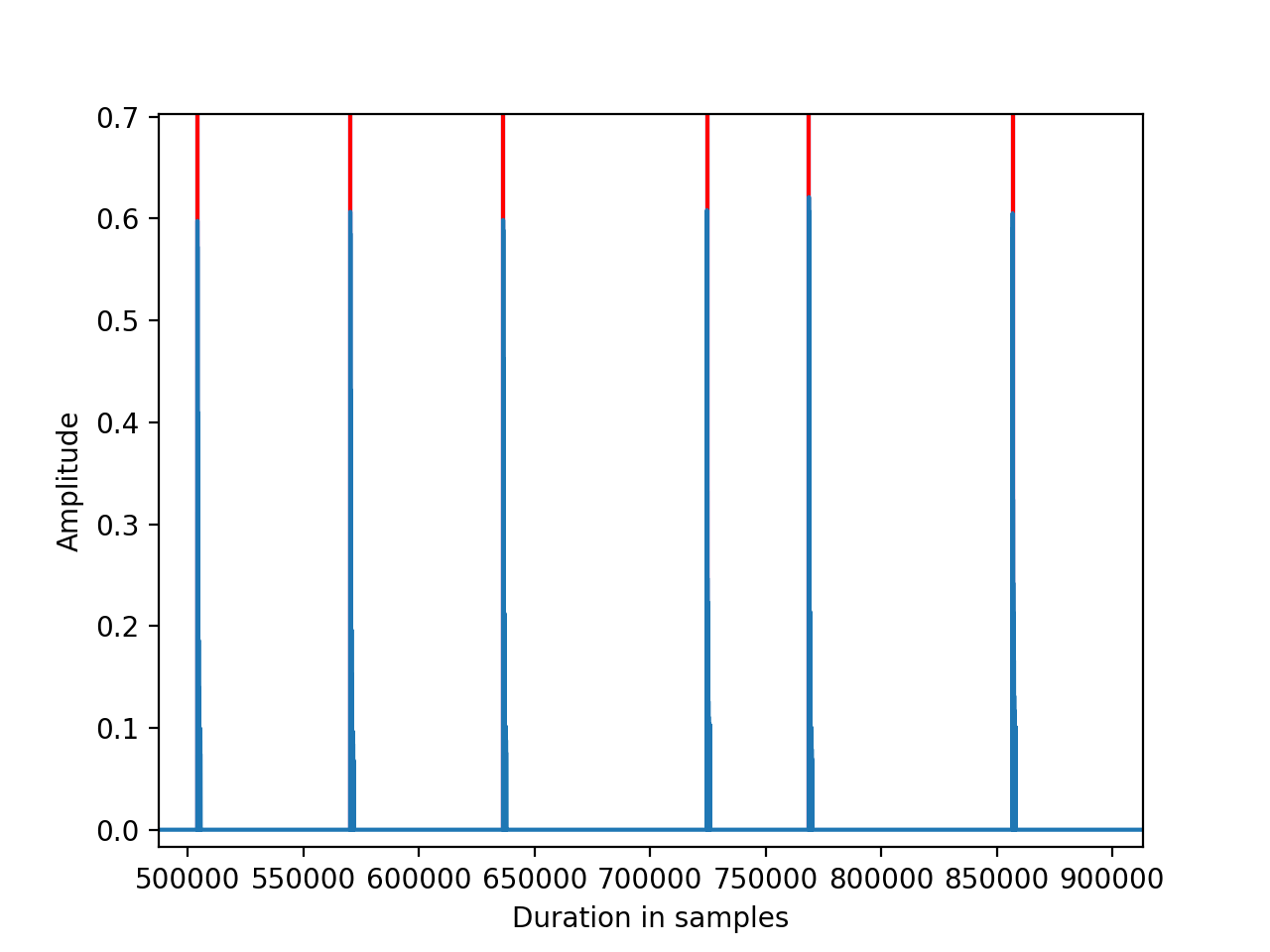}}
			\centerline{(b) Onsets from rectified inputs.}\medskip
		\end{minipage}

		\caption{Plots of onset detection in one cycle.}
		\label{fig:onset}
		
	\end{figure}
	
	Any onset detection function can be used to extract the onsets from the rectified recorded input. We used the onset detection function defined by the Librosa python library [29]. A sample of the recorded input and the extracted onsets are shown in figure \ref{fig:onset}.
	
	Inter onset intervals are calculated from onsets estimated from both recorded ($IOI_{input}$) and generated signals ($IOI_{generated}$). \\The deviation: $IOI_{generated} - IOI_{input}$, for each inter onset interval is calculated to be the error in temporal performance at each onset.
	
\begin{figure}[]
		\centering
		\centerline{\includegraphics[width=5cm, height=1.3cm]{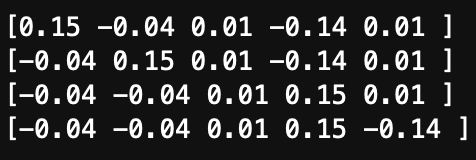}}
		\caption{$4X5$ Matrix of error deviation result.}
		\label{fig:matrix1}
	\end{figure}

	The error is recorded in a matrix format where the number of columns is equivalent to the number of onsets in a cycle and the number of rows is determined by the total number of cycles. For the son clave rhythm, there are 5 \enquote*{1's} (onsets) and 11 \enquote*{0's} (silence) played for 4 cycles, resulting in a $4X5$ matrix as shown in \ref{fig:matrix1}.
	
\begin{figure}[h!]
		\centering
		\centerline{\includegraphics[width=8.75cm,height = 5cm]{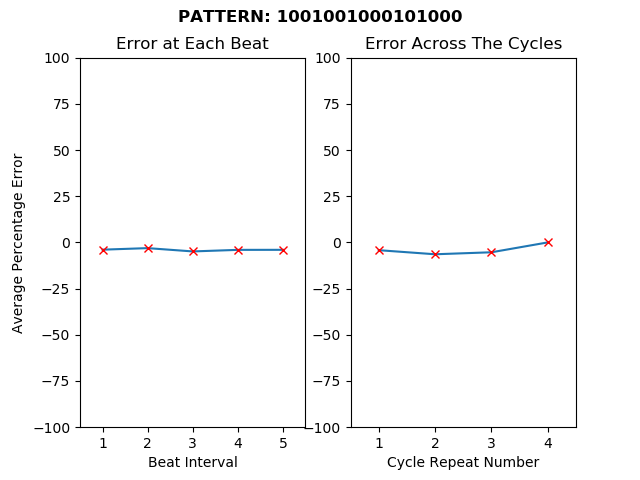}}
		\caption{Deviation graph results of the performance assessment block.}
		\label{fig:res1}
	\end{figure}

	From the error matrix, two sets of averages are calculated. One set of average values represent the deviation of each inter onset interval within a cycle (average of all the values in one column of the error matrix). This results in a $5X1$ average onset error array. This average onset error array is plotted on a graph with the horizontal axis representing each onset and the vertical axis representing the percentage of deviation (-100 to 100) This is shown in figure \ref{fig:res1}(left). This result indicates to the learner their timing error at each onset. The negative error implies slowly playing speed and the positive indicates fast playing speed. 
	\\
	The next set of average values represent the error deviation of a learner across the cycles (average all the values in one row of the error matrix) which results in a $4X1$ average result array.  This result is plotted on a graph with the horizontal axis representing the number of cycles and the vertical axis representing the deviation (-100 to 100) shown in figure \ref{fig:res1}(right). This result gives the learner information on their consistency throughout the performance and indicates if the timing error is consistent across cycles.

	The graphs in figure \ref{fig:res1} give the learner comprehensive performance feedback concerning their timing and consistency. We achieve this through the graphical representation wherein, a positive deviation (above 0 line) in figure \ref{fig:res1} (left) infers that the learner is rushing or playing the beat faster than intended and a negative deviation (below 0 line) infers that the learner is dragging or playing the beat slower than intended. Each onset in figure \ref{fig:res1} (left) is marked by an \textbf{x}.


	\section{1e0a - The Web Application}
	For improved accessibility to this system to aid in learning, we developed a web application hosted at \url{https://1e0a.noelalben.com}
	
	Our goal was to implement the proposed system with a simple user interface and a server to collect learners' data. We were able to achieve both by using Flask a micro web framework written in Python [30]. The website was hosted on AWS [31] with certbot SSL certification [32].\\
	
	\begin{figure}[h!]

		\begin{minipage}[b]{0.48\linewidth}
			\centering
			\centerline{\includegraphics[width=4.0cm]{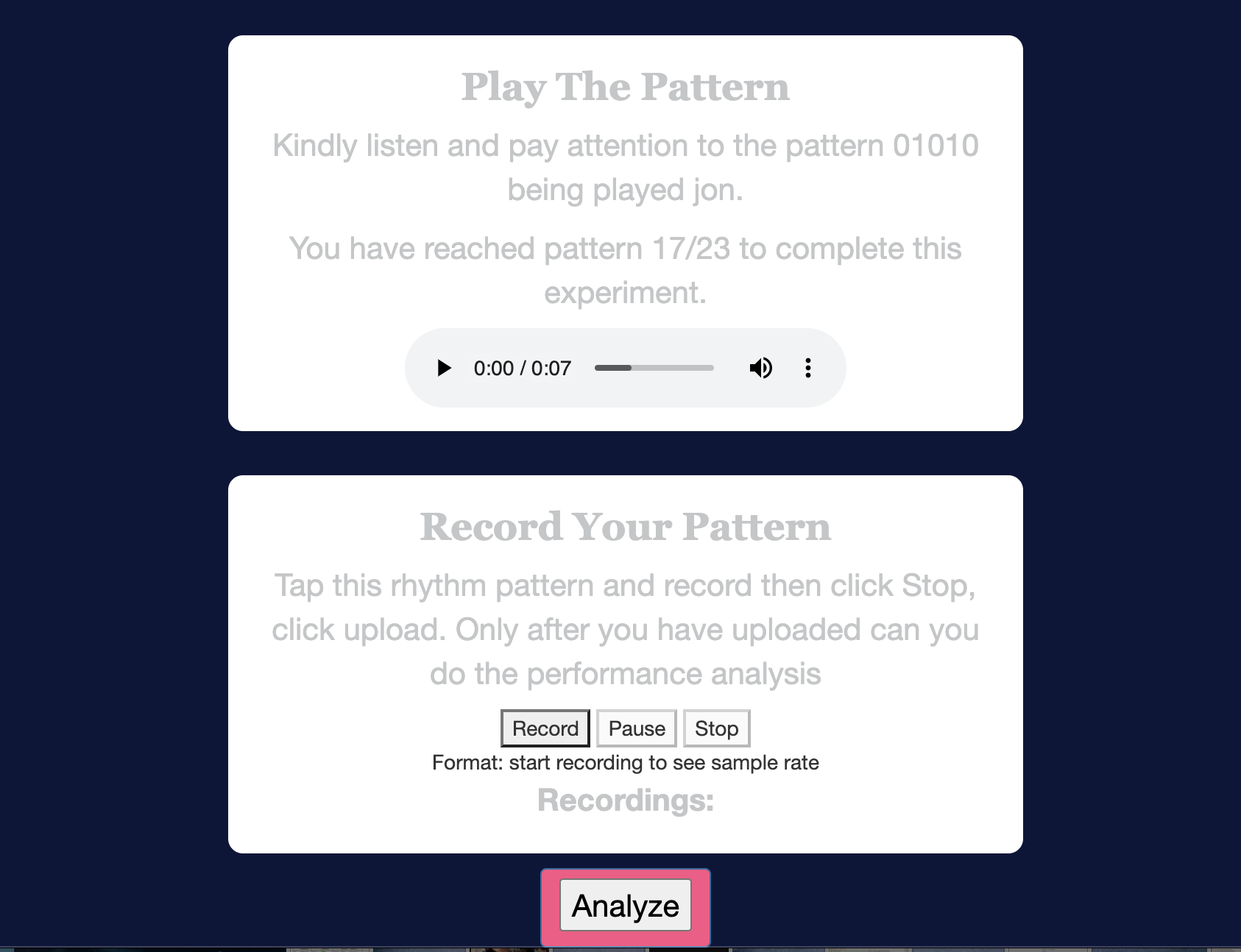}}
			\centerline{(a) Learning platform}\medskip
		\end{minipage}
		\hfill
		\begin{minipage}[b]{0.48\linewidth}
			\centering
			\centerline{\includegraphics[width=4.0cm]{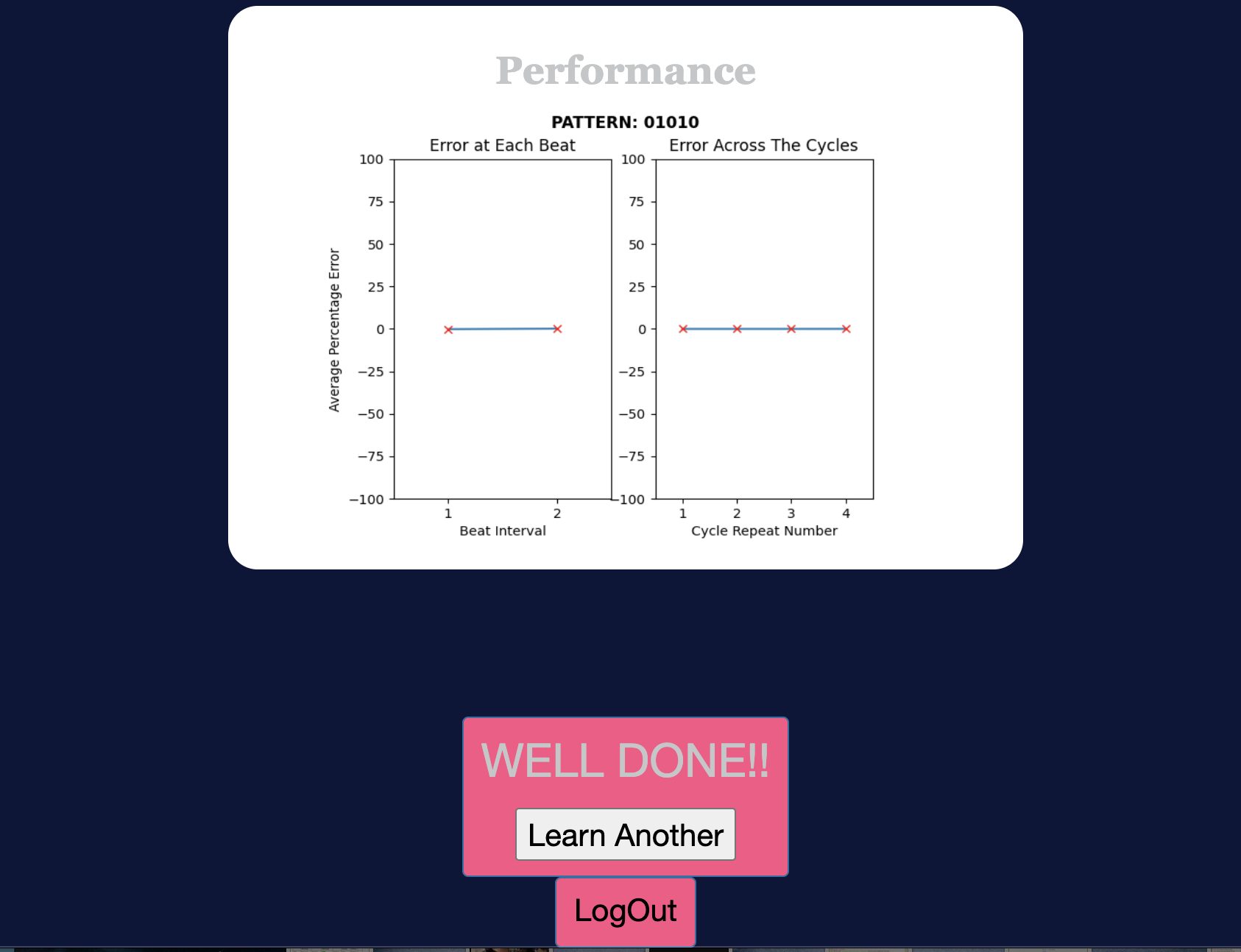}}
			\centerline{(b) Performance Assessment results}\medskip
		\end{minipage}

		\caption{\enquote*{1e0a} Web Application.}
		\label{fig:web1}
		
	\end{figure}
	Every new learner must register with the website and provide consent to use their registration information and performance assessment results for research purposes. Post-registration the learner is given a login id and password.
	Upon logging in to the system they enter the learning platform as shown in figure \ref{fig:web1} (a). They must listen to the generated pattern and record their respective performance. The \enquote*{Analyze} button conducts performance assessment on the learner's recording and takes them to the results as shown in figure \ref{fig:web1} (b). If they choose to progress in their learning by clicking \enquote*{Learn Another}, the system will generate the next pattern based on the complexity metrics defined in section \ref{complexity} and take the learner back to the learning platform. When they log out, the most recent generated pattern is stored into the database and they can continue where they left, upon logging in again.

	\begin{figure*}[h]
		\begin{minipage}[b]{0.2\linewidth}
			
			{\includegraphics[width=4cm]{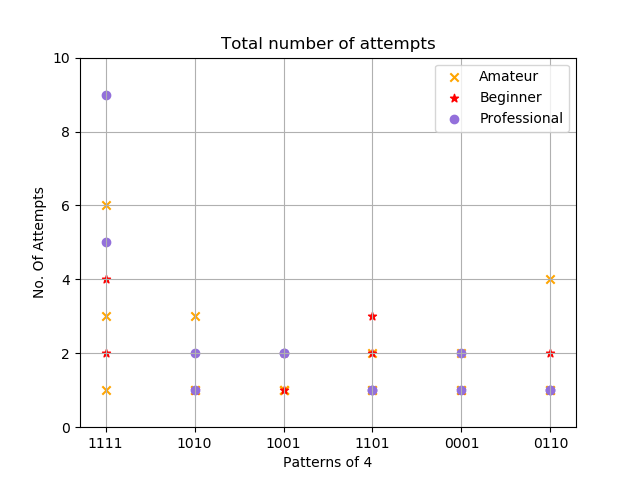}}
			\centerline{4(a)}
		\end{minipage}
		\begin{minipage}[b]{0.2\linewidth}
			\hfill
			
			{\includegraphics[width=4cm]{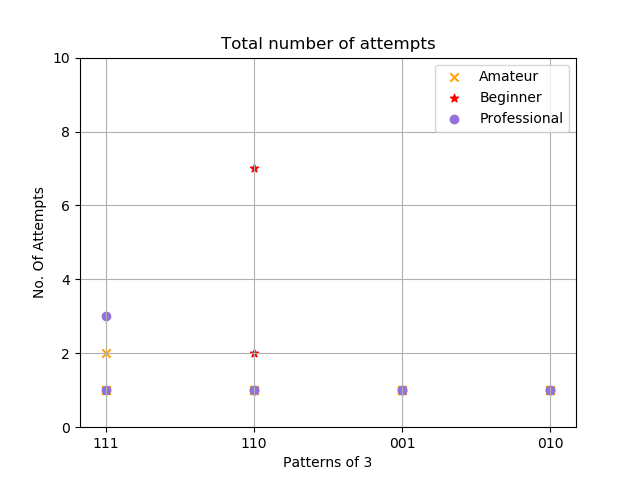}}
			\centerline{ 3(a) }\hfill
		\end{minipage}
		\begin{minipage}[b]{0.2\linewidth}
			
			{\includegraphics[width=4cm]{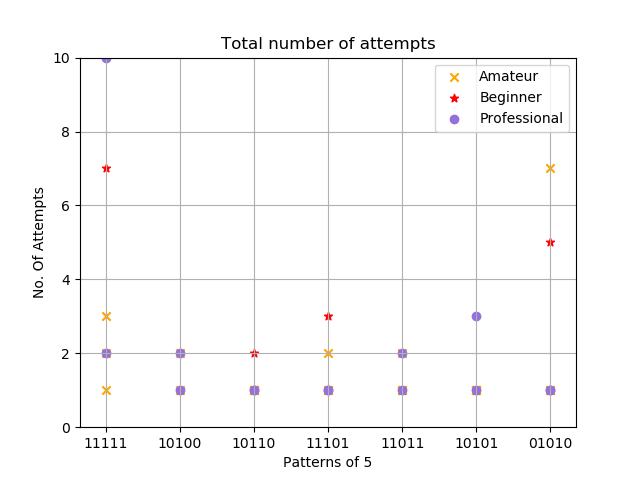}}
			\centerline{ 5(a)}
		\end{minipage}
		\begin{minipage}[b]{0.2\linewidth}
			\hfill
			
			{\includegraphics[width=4cm]{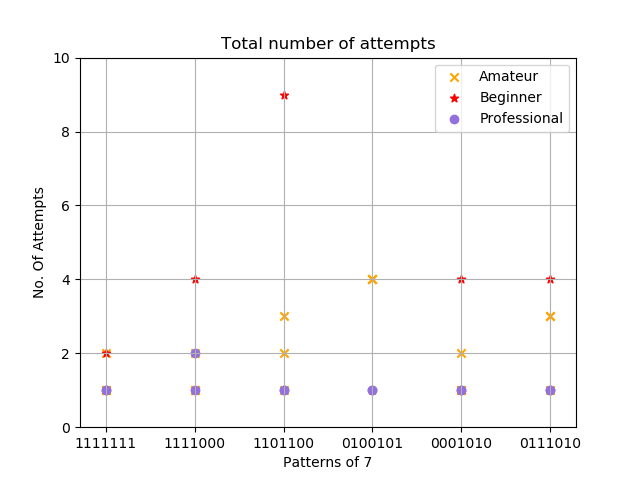}}
			\centerline{ 7(a) }
		\end{minipage}
		\begin{minipage}[b]{0.2\linewidth}
			
			{\includegraphics[width=4cm]{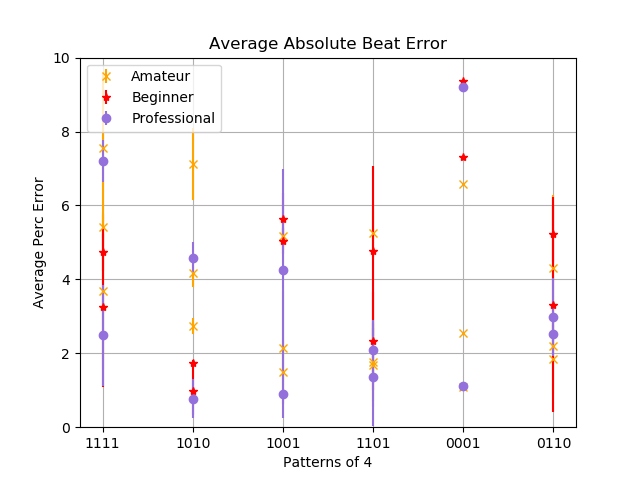}}
			\centerline{ 4(b)}
		\end{minipage}
		\begin{minipage}[b]{0.2\linewidth}
			\hfill
			
			{\includegraphics[width=4cm]{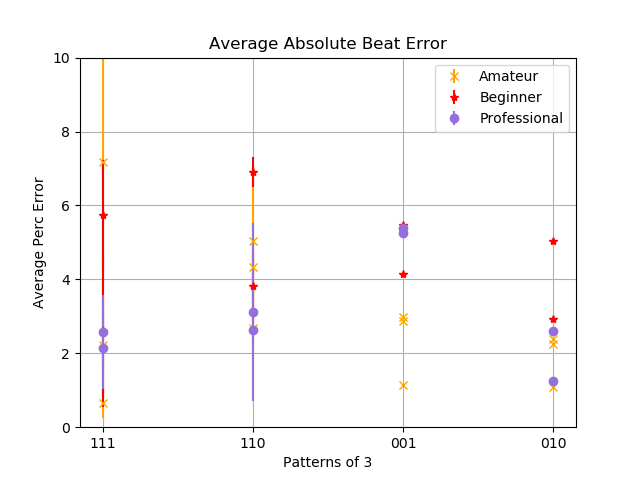}}
			\centerline{ 3(b) }
		\end{minipage}
		\begin{minipage}[b]{0.2\linewidth}
			
			{\includegraphics[width=4cm]{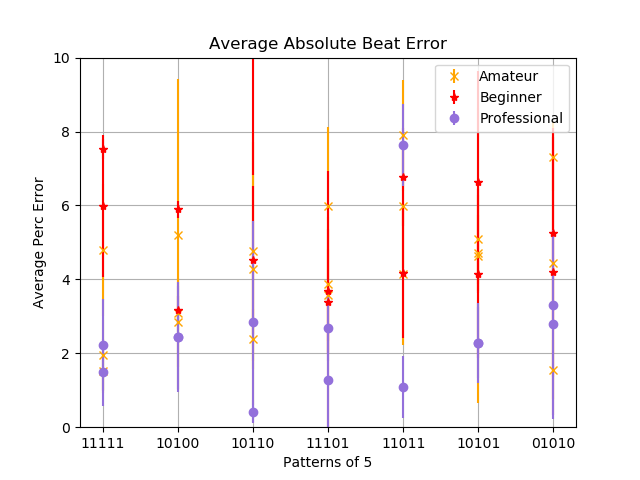}}
			\centerline{ 5(b)}
		\end{minipage}
		\begin{minipage}[b]{0.2\linewidth}
			\hfill
			
			{\includegraphics[width=4cm]{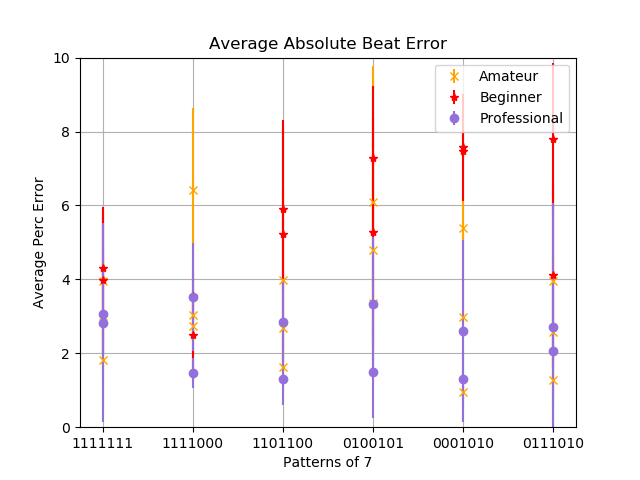}}
			\centerline{ 7(b) }
		\end{minipage}
		\begin{minipage}[b]{0.2\linewidth}
			
			{\includegraphics[width=4cm]{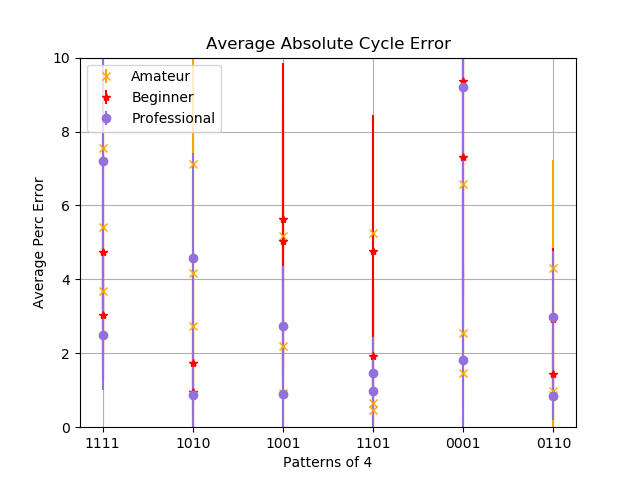}}
			\centerline{ 4 (c)}
		\end{minipage}
		\begin{minipage}[b]{0.2\linewidth}
			\hfill
			
			{\includegraphics[width=4cm]{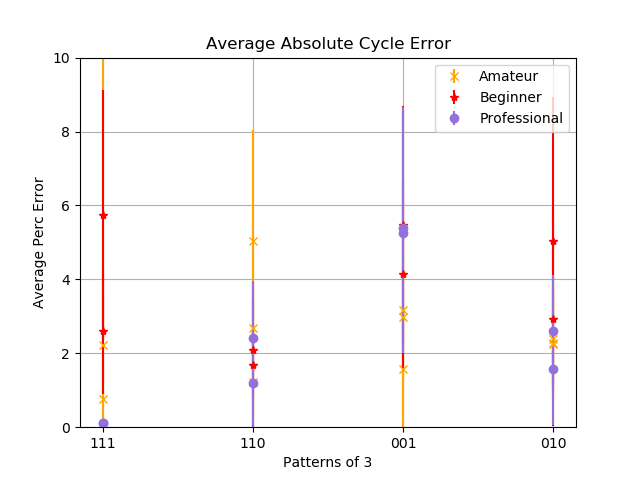}}
			\centerline{ 3(c) }
		\end{minipage}
		\begin{minipage}[b]{0.2\linewidth}
			
			{\includegraphics[width=4cm]{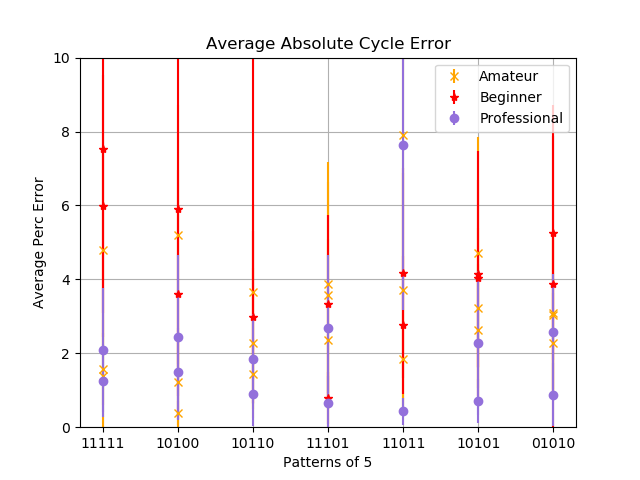}}
			\centerline{ 5(c)}
		\end{minipage}
		\begin{minipage}[b]{0.2\linewidth}
			\hfill
			
			{\includegraphics[width=4cm]{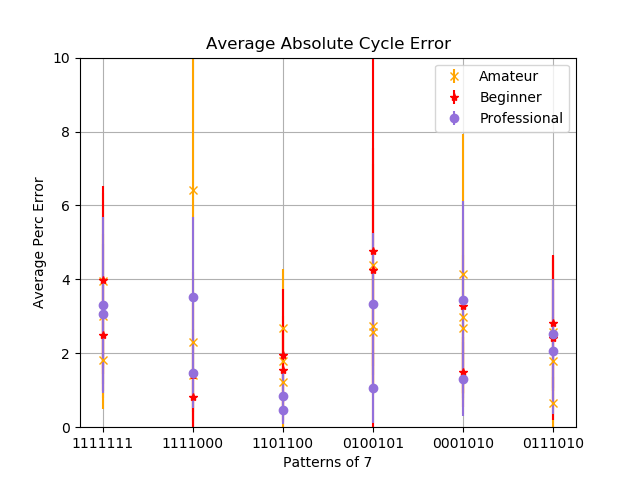}}
			\centerline{ 7(c) }
		\end{minipage}
		\begin{minipage}[b]{0.2\linewidth}
			\hfill
		\end{minipage}
		
		\caption{Graphical analysis of performance assessment for 6 learners categorised by professional($\bullet$), amateur($\times$), and beginner($\star$) on the 1e0a web application at each cycle. (a)Top Panel: Total number of attempts taken by learners.
			(b)	Center Panel: Absolute average beat error in performance.
			(c) Bottom Panel:  absolute average error over the cycle in performance. }
		\label{fig:web12}
	\end{figure*}
	\section{Experiments and Results}
	\subsection{Experimental Setup}
	To assess and quantify the effectiveness of our rhythm learning system, each learner on the 1e0a web application was required to perform and record 23 distinct patterns over 4 different cycle lengths of (4,3,5 and 7), at a constant PPM of 160. The learners were asked to use a desktop/laptop to access the web application and to maintain a constant interface to perform the rhythmic patterns (clapping or a single percussive instrument). They were not allowed to use accents in their performance nor take breaks while performing a given pattern. Throughout the experiment, the application recorded the number of attempts each learner takes to achieve performance assessment results within the error bounds presented in section \ref{assess}. Furthermore, we stored the performance assessment results of the learner for each rhythmic pattern where their results were within the error bounds. The results of 6 learners are shown in fig \ref{fig:web12}. Each learner is categorized based on their years of experience with music education: Professionals (>10 years), Amateur (>2
	years), and Beginners( 0-2 years).
	\subsection{Results}
	From the resulting graphs in figure \ref{fig:web12}, it can be observed that compared to an amateur or a beginner, for any given pattern the professional requires less number of attempts to achieve performance accuracy within the error bounds. This is reflective of a learning system where beginners make more mistakes and progressively improve in their performance as they spend more time on the application. We also observe that the learners were able to correct their mistakes and achieve performance accuracy within the error bounds after a sufficient number of attempts. This indicates that the performance assessment feedback of the system aids the learner in understanding and correcting their performance to achieve accuracy. Furthermore, it can be observed that across different patterns within the cycle there exists complexity because the number of attempts taken by the learner increases with the progression of patterns within a cycle. However, this does not directly reflect the difficulty that learners have while performing or listening to the rhythm; but does accurately reflect how beginners and amateurs struggle to recognize a rhythm's structure as the complexity increases.

	\section{Conclusion}
	In this article, we introduce \enquote*{1e0a}, the proof of concept for a rhythm learning and assessment system that helps learners progress in their learning with a new pattern generated based on a pre-defined rhythm complexity measure. We describe the rhythm complexity measure that was used and present the methodology for rhythm performance assessment. The results of the proposed system show that we can successfully test a learner's proficiency on rhythm patterns over a single PPM and monitor their progression within a cycle of patterns of increasing complexity values. This system does not test the learner's proficiency across different tempos neither does it test for mixed cycle complexity. \\
	The future work is to implement a complexity measure that extends to various tempos, accents and mixed cycle length progression. Furthermore, we aim to have an improved onset detection and noise removal on the recorded audio signal for a performance assessment that is robust to noise and disturbances.





\end{document}